\documentclass[12pt,a4paper]{article}

\usepackage{amsfonts,amsmath,amssymb,amsthm}
\usepackage{latexsym,amscd}
\usepackage{amsbsy}
\usepackage{CJK}
\usepackage{indentfirst,mathrsfs}
\usepackage{amsfonts,amsmath,amssymb,amsthm}
\usepackage{latexsym,amscd}
\usepackage{amsbsy}
\usepackage{color}
\usepackage{multirow}
\usepackage{makecell}
\usepackage{float}
\usepackage{cases}

\numberwithin{equation}{section}

\newtheorem{thm}{Theorem}

\newtheorem{remark}{Remark}
\newtheorem{conj}{Conjecture}
\newcommand{\Tr}{{\rm {Tr}}}

\begin{document}

\title{On the differential spectrum of a class of power functions over finite fields}
\author{Nian Li, Yanan Wu, Xiangyong Zeng
        and Xiaohu Tang
\thanks{The authors are with the Hubei Key Laboratory of Applied Mathematics, Faculty of Mathematics and Statistics, Hubei
University, Wuhan, 430062, China. Xiaohu Tang is also with the Information Security and National Computing Grid Laboratory, Southwest Jiaotong University, Chengdu, 610031, China. Email: nian.li@hubu.edu.cn, yanan.wu@aliyun.com, xzeng@hubu.edu.cn, xhutang@swjtu.edu.cn}
}
\date{}
\maketitle

\begin{quote}
{\small {\bf Abstract:}
Differential uniformity is a significant concept in cryptography as it quantifies the degree of security of S-boxes respect to differential attacks. Power functions of the form $F(x)=x^d$ with low differential uniformity have been extensively studied in the past decades due to their strong resistance to differential attacks and low implementation cost in hardware.
In this paper, we give an affirmative answer to a recent conjecture proposed by Budaghyan, Calderini, Carlet, Davidova and Kaleyski about the
differential uniformity of $F(x)=x^d$ over $\mathbb{F}_{2^{4n}}$, where $n$ is a positive integer and $d=2^{3n}+2^{2n}+2^{n}-1$, and we completely determine its differential spectrum.
}

\end{quote}

\section{Introduction} \label{intro}

Let $n$, $m$ be two positive integers and $\mathbb{F}_{2^n}$ denote the finite field with $2^n$ elements. An S-box is a vectorial Boolean function from  $\mathbb{F}_{2^n}$ to $\mathbb{F}_{2^m}$, also called an $(n,m)$-function. The security of most modern block ciphers importantly relies on cryptographic properties of their S-boxes since S-boxes usually are the only nonlinear elements of these cryptosystems. It is therefore significant to employ S-boxes with good cryptographic properties in order to resist various kinds of cryptanalytic attacks.

Differential attack \cite{BS} is one of the most fundamental cryptanalytic approaches targeting symmetric-key primitives and is the first statistical attack for breaking iterated block ciphers. The differential uniformity of S-boxes, which was introduced by Nyberg in \cite{N}, can be used to measure how well the S-box used in the cipher could resist the differential attack.
An $(n,n)$-function $F$ is called $\delta_F$-differential uniform if the equation
$F(x+a)+F(x)=b$ has at most $\delta_F$ solutions for any $a\in\mathbb{F}_{2^n}^*$ and $b\in\mathbb{F}_{2^n}$, i.e.,
$$\delta_F=\max \{\delta(a, b):a\in\mathbb{F}_{2^n}^*,\,b \in \mathbb{F}_{2^n} \}$$
where
\begin{eqnarray}\label{eq-delta}
  \delta(a, b)=\#\left\{x \in \mathbb{F}_{2^{n}}: F(x+a)+F(x)=b\right\}.
\end{eqnarray}
Observe that if $x$ is a solution to $F(x+a)+F(x)=b$ then so is $x+a$. This shows that $\delta(a,b)$ is even and $\delta_F\geq 2$. The $(n,n)$-function $F$ with $\delta_F=2$ is called almost perfect nonlinear (APN) and provides the best possible resistance to differential attacks.

Power functions, namely, monomial functions, as a special class of functions over finite fields, have been extensively studied in the last decades due to their simple algebraic form and lower implementation cost in hardware environment. Very recently, Budaghyan, Calderini, Carlet, Davidova and Kaleyski in \cite{BCCDK} presented some observations and computational data on the differential spectra of power functions $F(x)=x^d$ with $d=\sum_{i=1}^{k-1}2^{in}-1$ over the finite field $\mathbb{F}_{2^{nk}}$, where $n, k$ are positive integers. It is worth noting that this class of power functions includes some famous functions as special cases. For example, if one takes $n=1$, then $F(x)=x^d$ is exactly the well-known inverse function  which is either APN or 4-differential uniform and has been widely used in practical cryptosystems. If $k=3$ or $k=5$, then $F(x)=x^d$ is the Kasami \cite{Dobbertin,Hollmann,Janwa,Kasami} or Dobbertin \cite{Dobbertin-apn} type of APN power function respectively, which are another two families of the six known infinite families of  APN power functions. The differential property of $F(x)=x^d$ has also been studied for $k=2$ in \cite{Anne, Lilya} while that of $F(x)=x^d$ for the case of $k=4$ remains unknown. This motivated the authors in \cite{BCCDK} and based on experimental data they proposed the following conjecture:

\begin{conj}{\rm(\cite[Conjectrue 22]{BCCDK})\label{conj}}
  Let $d=2^{3n}+2^{2n}+2^n-1$ and consider the power function $x^d$ over $\mathbb{F}_{2^{4n}}$. Then the equation $x^d+(x+1)^d=b$
has $2^{2n}$ solutions for one value of $b$; it has $2^{2n}-2^n$ solutions for $2^n$ values of $b$; and has at most $2$ solutions for all remaining
points $b$.
\end{conj}

If Conjectrue \ref{conj} is settled, then the differential property of $F(x)=x^d$ for $k=4$ can be completely determined.
This paper aims to settle Conjectrue \ref{conj} and then determine the differential spectrum of $F(x)=x^d$ for $k=4$ by employing some particular techniques in solving the equation in Conjectrue \ref{conj}.

\section{Main Result}\label{main}

Let $F(x)=x^d$ be a power function on $\mathbb{F}_{2^{4n}}$, where $d=2^{3n}+2^{2n}+2^n-1$. Note that  $\delta(a, b)=\delta(1, b/a^d)$, where $\delta(a, b)$ is defined as \eqref{eq-delta}. Hence the differential characteristics of $F(x)=x^d$ are completely determined by the values of $\delta(1, b)$ for $b\in\mathbb{F}_{2^{4n}}$.  Let $\omega_i$ be defined as follows:
\begin{eqnarray*}
\omega_i=\#\{b\in\mathbb{F}_{2^{4n}}: \delta(1, b)=i\}.
\end{eqnarray*}
The differential spectrum of $F(x)=x^d$ is the set $\mathbb{S}$ of $\omega_i$ where $i$ is even and $0\leq i\leq \delta_F$ (since $\omega_i=0$ if $i$ is odd):
\begin{eqnarray*}
\mathbb{S}=\{\omega_0,\omega_2,\cdots, \omega_{\delta_F}\}.
\end{eqnarray*}

As was pointed out in \cite{Blondeau,Xiong}, it is of interest to obtain the differential spectrum of power functions with low differential uniformity which is useful to analyse the resistance of the cipher to differential attacks. Moreover, the problem of computing differential spectrum is also interesting and challenging from mathematical point of view.

Our main result is given as below.

\begin{thm}\label{thm}
Let $F(x)=x^d$ be a power function on $\mathbb{F}_{2^{4n}}$, where $d=2^{3n}+2^{2n}+2^n-1$. The differential spectrum of $F(x)$ is given by $\mathbb{S}=\{\omega_0, \omega_2, \omega_{2^{2n}-2^n}, \omega_{2^{2n}}\}$ where
\begin{eqnarray*}
\begin{array}{llllll}
  \omega_0 &=& (2^{3n-1}-1)(2^n+1),  &\omega_2 &=& 2^{4n-1}-2^{3n-1}, \\
  \omega_{2^{2n}-2^n}&=& 2^n, & \omega_{2^{2n}} &=& 1.
\end{array}
\end{eqnarray*}
\end{thm}

Theorem \ref{thm} not only gives an affirmative answer to Conjecture \ref{conj}  but also completely determines the differential spectrum of the power function $F(x)=x^d$ on $\mathbb{F}_{2^{4n}}$ for $d=2^{3n}+2^{2n}+2^n-1$.

\begin{remark}
 It is interesting to note that the exponent $d=2^{3n}+2^{2n}+2^n-1$ is a Niho exponent with respect to the finite field $\mathbb{F}_{2^{4n}}$. This exponent has been essentially (under the cyclotomic equivalence) studied in \cite{Dobbertin-it,Niho} for the cross correlation between an $m$-sequence and its $d$-decimation sequence in which the Walsh spectrum of the power function $F(x)=x^d$ has been determined.
\end{remark}

\section{Proof of the Main Result}\label{proof}

From now on, we always assume that $n$ is a positive integer, $q=2^n$ and $d=q^3+q^2+q-1$.
Let $s$ be a positive integer and define
\begin{eqnarray*}
  \mu_s=\{x\in\mathbb{F}_{q^{4}}: x^s =1\}.
\end{eqnarray*}
%

To complete the proof of Theorem \ref{thm}, it is sufficient to determine the number of solutions $x\in\mathbb{F}_{q^4}$ of
\begin{eqnarray}\label{diff-eq}
(x+1)^d+x^d=b
\end{eqnarray}
when $b$ runs through $\mathbb{F}_{q^4}$.
Note that
\begin{eqnarray}\label{eq-d}
d(q+1)\equiv 2q^2(q+1)\,({\rm mod}\,\, q^4-1).
\end{eqnarray}
Using \eqref{diff-eq} and \eqref{eq-d}, one obtains
  \begin{equation}\label{diff-q+1}
  (x^{2q^2}+1)^{q+1}=(x^d+b)^{q+1}.
  \end{equation}

Then we can discuss the solutions of \eqref{diff-eq} as follows:

Case 1: $b=0$.

In this case \eqref{diff-eq} has no solution due to $\gcd(d,q^4-1)=1$.

Case 2: $b=1$.

If $b=1$, then \eqref{diff-q+1} is reduced to $(x^d+x^{2q^3})(1+x^{dq-2q^3})=0$ which leads to $x^d=x^{2q^3}$ or $x^{dq-2q^3}=1$.
 For the former case,  one has  $x=0$ or $x^{(q^2-1)(1-q)}=1$, which implies that $x\in\mathbb{F}_{q^2}$ due to $\gcd((q^2-1)(q-1),q^4-1)=q^2-1$. It can be easily verified that any $x\in\mathbb{F}_{q^2}$ is a solution of \eqref{diff-eq} when $b=1$. For the latter case, one gets $x^{(q^2+1)(q-1)}=1$, i.e., $x\in\mu_{(q^2+1)(q-1)}$. Note that every $x\in\mu_{(q^2+1)(q-1)}$ can be uniquely expressed as $x=yz$ for some $y\in\mu_{q-1}$ and $z\in\mu_{q^2+1}$ due to $\gcd(q-1,q^2+1)=1$. Then \eqref{diff-eq} becomes
 $(yz+1)^d+(yz)^d=1$
 which can be written as
  \begin{eqnarray*}
 (yz+1)^{q^3}(yz+1)^{q^2}(yz+1)^{q}+y^dz^d(yz+1)=yz+1,
 \end{eqnarray*}
 i.e.,
 \begin{eqnarray*}
 (yz^{-q}+1)(yz^{-1}+1)(yz^q+1)+y^2z^{-2}(yz+1)=yz+1.
 \end{eqnarray*}
A straightforward calculation gives
  \begin{eqnarray*}
 (yz^{-1}+1)(z^{q+1}+1)(z^{-q}+z^{-1})=0.
 \end{eqnarray*}
This identity holds if and only if $z=1$ since $z\in\mu_{q^2+1}$ and then one has that $x=y\in\mu_{q-1}\subset \mathbb{F}_{q^2}$. Combining above discussions one can conclude that \eqref{diff-eq} has $q^2$ solutions when $b=1$.

Case 3: $b\not\in\mathbb{F}_2$.

Note that $x=0,1$ are solutions of \eqref{diff-eq} if and only if $b=1$.
For this case, according to \eqref{diff-q+1}, one has that there exists some $\gamma_1\in\mu_{q+1}$ such that
 \begin{equation}\label{diff-q+1-gamma1}
 x^d=b+\gamma_1(x^{2q^2}+1).
 \end{equation}
 Raising both sides of \eqref{diff-q+1-gamma1} to the $(q+1)$-th power gives
 $$(x^{2q^2})^{q+1}=(b+\gamma_1(x^{2q^2}+1))^{q+1}$$
due to \eqref{eq-d}. This implies that there exists some $\gamma_2\in\mu_{q+1}$ such that
 \begin{equation}\label{diff-q+1-gamma2}
 \gamma_2x^{2q^2}=b+\gamma_1(x^{2q^2}+1),\;\;{\rm i.e.},\;\; (\gamma_1+\gamma_2)x^{2q^2}=b+\gamma_1.
 \end{equation}

Case 3.1:  $\gamma_1=\gamma_2$.

If this case occurs,  one then obtains $b=\gamma_1\in\mu_{q+1}\backslash\{1\}$ and $x^d=bx^{2q^2}$ from  \eqref{diff-q+1-gamma2} and \eqref{diff-q+1-gamma1} respectively. Thus in this case \eqref{diff-eq} can be written as
 \begin{equation*}
\left\{\begin{array}{cll}
x^d&=&bx^{2q^2},\\
(x+1)^d&=&b(x+1)^{2q^2}.
\end{array}\right.
\end{equation*}
Observe that $d-2q^2=(q^2+1)(q-1)$ and then the above system of equations becomes
 \begin{equation*}
\left\{\begin{array}{cll}
x^{(q^2+1)q}&=&bx^{q^2+1},\\
(x+1)^{(q^2+1)q}&=&b(x+1)^{q^2+1}
\end{array}\right.
\end{equation*}
which can be further reduced to
\begin{eqnarray}\label{eq-b=gamma1}
\left\{\begin{array}{cll}
x^{(q^2+1)q}&=&bx^{q^2+1},\\
(x^{q^2}+x+1)^q&=&b(x^{q^2}+x+1).
\end{array}\right.
\end{eqnarray}

The number of solutions of \eqref{eq-b=gamma1} can be determined as follows:

(1) Notice that $x^{q^2}+x+1\in\mathbb{F}_{q^2}$ and in this case we have $1\ne b\in\mu_{q+1}$. Let $b=w^{q-1}$ for some $w\in\mathbb{F}_{q^2}\backslash\mathbb{F}_q$ and $y=x^{q^2}+x+1$. Then by the second equation in \eqref{eq-b=gamma1} one obtains $y=0$ or $y^{q-1}=b=w^{q-1}$ which implies that there exists some $\sigma\in\mathbb{F}_q$ such that $y=\sigma w$. For a fixed $\sigma\in \mathbb{F}_q$, suppose that $x_0$ is a solution of $x^{q^2}+x+1+\sigma w=0$, i.e., $x_0^{q^2}=x_0+1+\sigma w$, then $x_0$ is a solution of \eqref{eq-b=gamma1} if $x_0^{(q^2+1)q}=bx_0^{q^2+1}$. That is, $x_0^q(x_0+1+\sigma w)^q=w^{q-1}x_0(x_0+1+\sigma w)$ which can be rewritten as $(x_0(x_0+1+\sigma w))^{q-1}=w^{q-1}$. This indicates that there exists some $\epsilon\in\mathbb{F}_{q}^*$ such that
\begin{equation}\label{quadratic-eq-1}
x_0^2+(1+\sigma w)x_0=\epsilon w.
\end{equation}
Note that $1+\sigma w\ne 0$ due to $\sigma\in\mathbb{F}_q$ and $w\in\mathbb{F}_{q^2}\backslash\mathbb{F}_q$. Let $x_0=(1+\sigma w)z$, then \eqref{quadratic-eq-1} becomes
 \begin{equation}\label{quadratic-eq-2}
 z^2+z+\frac{\epsilon w}{(1+\sigma w)^2}=0.
 \end{equation}
Again by the facts $\sigma\in\mathbb{F}_q$ and $w\in\mathbb{F}_{q^2}\backslash\mathbb{F}_q$ one has that \eqref{quadratic-eq-2} has two solutions in either $\mathbb{F}_{q^2}$ or $\mathbb{F}_{q^4}$. On the other hand, it can be readily verified that any $x\in\mathbb{F}_{q^2}$ is not a solution of \eqref{eq-b=gamma1} if $b\ne 0,1$. Thus one can claim that \eqref{quadratic-eq-2} has no solution in $\mathbb{F}_{q^2}$ which implies that the elements $\sigma\in\mathbb{F}_q$ and $\epsilon\in\mathbb{F}_q^*$ should satisfy
\begin{eqnarray}\label{eq-tr}
 \Tr^{2n}_{1}(\frac{\epsilon w}{(1+\sigma w)^{2}})=\Tr^{ n}_{1}(\epsilon (\frac{w}{(1+\sigma w)^{2}}+\frac{w^q}{(1+\sigma w)^{2q}}))=1,
\end{eqnarray}
where $\Tr^{n}_{1}(\cdot)$ is the trace function from $\mathbb{F}_{2^{n}}$ to $\mathbb{F}_{2}$. A direct calculation shows
$$\frac{w}{(1+\sigma w)^{2}}+\frac{w^q}{(1+\sigma w)^{2q}}=0$$ if and only if $\sigma^2w^{q+1}=1$, i.e., $\sigma=w^{-\frac{1}{2}(q+1)}$. Hence, if $\sigma\not=w^{-\frac{1}{2}(q+1)}$, then there are $\frac{q}{2}$ choices of $\epsilon$ satisfying \eqref{eq-tr}. Further, for each such pair $(\sigma, \epsilon)$, \eqref{quadratic-eq-2} has two solutions in $\mathbb{F}_{q^4}\backslash\mathbb{F}_{q^2}$, and so does \eqref{quadratic-eq-1}.

(2) Next we show that the two solutions of \eqref{quadratic-eq-1} obtained from any $(\sigma, \epsilon)$ satisfying \eqref{eq-tr} are solutions of \eqref{eq-b=gamma1}. Raising \eqref{quadratic-eq-2} to the $2^i$-th power for $i=0,\,1,\,\cdots,2n-1$ and adding them together gives
 \begin{eqnarray*}
z^{q^2}+z=\Tr^{2n}_{1}(\frac{\epsilon w}{(1+\sigma w)^{2}})=1.
 \end{eqnarray*}
Then by $x_0=(1+\sigma w)z$, one gets
 \begin{eqnarray*}
x_0^{q^2}+x_0+1+\sigma w=(1+\sigma w)(z^{q^2}+z+1)=0
 \end{eqnarray*}
 which together with \eqref{quadratic-eq-1} implies that
  \begin{eqnarray*}
x_0^{(q^2+1)q}=bx_0^{q^2+1},
 \end{eqnarray*}
 i.e., $x_0$ is a solution of \eqref{eq-b=gamma1}.

 (3) Thirdly we show that any different pair $(\sigma, \epsilon)$ gives distinct solutions of \eqref{quadratic-eq-1}. Let
 $(\sigma_1, \epsilon_1)$ and $(\sigma_2, \epsilon_2)$ satisfy \eqref{eq-tr} with $(\sigma_1, \epsilon_1)\ne(\sigma_2, \epsilon_2)$, then both $x^2+(1+\sigma_1 w)x=\epsilon_1 w$ and $x^2+(1+\sigma_2 w)x=\epsilon_2 w$ have no solution in $\mathbb{F}_{q^2}$. Note that the solutions of
  \begin{equation*}
\left\{\begin{array}{cll}
x^2+(1+\sigma_1 w)x=\epsilon_1 w,\\x^2+(1+\sigma_2 w)x=\epsilon_2 w
\end{array}\right.
\end{equation*}
 satisfies $$(\sigma_1+\sigma_2)x=(\epsilon_1+\epsilon_2)$$
 for $\sigma_i\in\mathbb{F}_q$ and $\epsilon_i\in\mathbb{F}_q^*$, where $i=1,2$. Since $\sigma_1+\sigma_2\ne 0$ otherwise we have $\epsilon_1+\epsilon_2=0$, one can conclude that the common solutions of $x^2+(1+\sigma_1 w)x=\epsilon_1 w$ and $x^2+(1+\sigma_2 w)x=\epsilon_2 w$ lie in $\mathbb{F}_q$ which is impossible.

Hence, according to the above discussions, we can claim that the number of solutions of \eqref{eq-b=gamma1} is $(q-1)\cdot\frac{q}{2}\cdot 2=q^2-q$  when $\gamma_1=\gamma_2$ and $b=\gamma_1\in\mu_{q+1}\backslash\{1\}$.

Case 3.2:  $\gamma_1\not=\gamma_2$.

If this case happens, then by \eqref{diff-q+1-gamma2} one obtains
\begin{equation}\label{eq-x}
x^{2q^2}=\frac{b+\gamma_1}{\gamma_1+\gamma_2}
\end{equation}
which is a solution of \eqref{diff-eq} if and only if
 \begin{equation}\label{gamma1 ne gamma2}
 \left(\frac{\gamma_1+b}{\gamma_1+\gamma_2}\right)^d+\left(\frac{\gamma_2+b}{\gamma_1+\gamma_2}\right)^d=b^{2q^2}.
 \end{equation}
Notice that $\gamma_i\in\mu_{q+1}$ for $i=1,2$ which implies $\gamma_i^q=\gamma_i^{q^3}=\gamma_i^{-1}$ and $\gamma_i^{q^2}=\gamma_i$. Then by a detailed calculation one can obtain that
 \begin{eqnarray*}
 (\gamma_i+b)^d=\frac{\gamma_ib^{q^3+q}+\gamma_i^{-1}(b^{q^3+q^2}+b^{q^2+q}+1)+\gamma_i^{-2}b^{q^2}+b^{q^3+q^2+q}+b^{q^3}+b^q}{\gamma_i+b}
\end{eqnarray*}
for any $b\ne \gamma_i$ and from which one can also have
 \begin{eqnarray*}
 (\gamma_1+\gamma_2)^d=\frac{\gamma_1^2+\gamma_2^2}{\gamma_1^2\gamma_2^2}.
\end{eqnarray*}
For simplicity, denote
\begin{eqnarray}\label{eq-AB}
A=b^{q^2+1}+b^{2q^2+2},\;\; B=b^{q^3+q^2+q}+b^{q^3+q+1}+b^{q^3}+b^q.
\end{eqnarray}
Then \eqref{gamma1 ne gamma2} can be rewritten as
 $$(\gamma_1+\gamma_2)B+(\frac{1}{\gamma_1}+\frac{1}{\gamma_2})B^q+(\frac{\gamma_2}{\gamma_1}+\frac{\gamma_1}{\gamma_2})(\frac{B^q}{b}+\frac{A}{b^2})
+\frac{(\gamma_1+\gamma_2)^3}{\gamma_1^2\gamma_2^2}\frac{A}{b}
 +(\frac{1}{\gamma_1^2}+\frac{1}{\gamma_2^2})A=0.$$
Multiplying $(\gamma_1+\gamma_2)^{-1}$ on both sides of the above equation gives
 \begin{equation}\label{gamma1 ne gamma2-1}
 B+\frac{1}{\gamma_1\gamma_2}B^q+\frac{\gamma_1+\gamma_2}{\gamma_1\gamma_2}\Big(\frac{B^q}{b}+\frac{A}{b^2}\Big)+
 \frac{\gamma_{1}^2+\gamma_{2}^2}{\gamma_{1}^2\gamma_{2}^2}\frac{A}{b}+ \frac{\gamma_{1}+\gamma_{2}}{\gamma_{1}^2\gamma_{2}^2}A=0.
  \end{equation}
  Taking $q^2$-th power on both sides of \eqref{gamma1 ne gamma2-1} and adding these two equations leads to
 \begin{eqnarray}\label{C}
B^qC+AC^2+\frac{\gamma_1+\gamma_2}{\gamma_1\gamma_2}AC=0
\end{eqnarray}
due to $A, B\in\mathbb{F}_{q^2}$ and $\gamma_i^{q^2}=\gamma_i$ for $i=1,2$, where $C=b^{-1}+b^{-q^2}$.

(1) If $C=0$, then $b\in\mathbb{F}_{q^2}\backslash\mathbb{F}_2$. This leads to $B=0$ and $A=b^4+b^2$ according to \eqref{eq-AB}. Then by \eqref{gamma1 ne gamma2-1} one has that
\begin{eqnarray*}
 \frac{\gamma_1+\gamma_2}{\gamma_1\gamma_2}\cdot(b^2+1)\cdot(1+\frac{\gamma_1+\gamma_2}{\gamma_1\gamma_2}b+\frac{b^2}{\gamma_1\gamma_2})=0
\end{eqnarray*}
which gives $(b^{-1}+\gamma_1^{-1})(b^{-1}+\gamma_2^{-1})=0$, i.e., $b=\gamma_1$ or $b=\gamma_2$. This leads  to $b=1$ by \eqref{gamma1 ne gamma2}, a contradiction. Thus in this case \eqref{gamma1 ne gamma2} has no solution for $\gamma_1,\gamma_2$ and then \eqref{diff-eq} has no solution in $\mathbb{F}_{q^4}$.

(2) If $C\ne0$, then $b\notin\mathbb{F}_{q^2}$ and by \eqref{C} one gets
$$\frac{\gamma_1+\gamma_2}{\gamma_1\gamma_2}A=B^q+AC.$$
If $A=0$, i.e., $b^{q^2+1}+b^{2(q^2+1)}=0$, one then has $b^{q^2+1}=1$ and $B^q=0$ which leads to $b+b^q+b^{q^2}+b^{q^3}=0$ by \eqref{eq-AB}. This together with $b^{q^2+1}=1$ implies that $b+b^{-1}\in\mathbb{F}_q$ which contradicts with $b\notin\mathbb{F}_{q^2}$. Thus we have $A\ne 0$ and
\begin{equation}\label{eq-gammasum}
\frac{\gamma_1+\gamma_2}{\gamma_1\gamma_2}=\frac{B^q}{A}+C=\frac{\Tr^{4n}_n(b)}{b^{q^2+1}+1}.
\end{equation}
Note that $$\frac{\gamma_1+\gamma_2}{\gamma_1\gamma_2}=\frac{1}{\gamma_1}+\frac{1}{\gamma_2}=\gamma_1^q+\gamma_2^q.$$
Then by \eqref{eq-gammasum} one can obtain
\begin{eqnarray*}
\left\{\begin{array}{cll}
\gamma_1\gamma_2&=&(b^{q^2+1}+1)^{1-q},\\
\gamma_1+\gamma_2&=&\Tr^{4n}_n(b)(b^{q^2+1}+1)^{-q},
\end{array}\right.
\end{eqnarray*}
i.e., $\gamma_1, \gamma_2$ are two solutions of
\begin{eqnarray}\label{eq-2nd}
  x^2+ \Tr^{4n}_n(b)(b^{q^2+1}+1)^{-q}x+(b^{q^2+1}+1)^{1-q}=0.
\end{eqnarray}
Moreover, one can claim that \eqref{eq-2nd} has either no solution or two solutions in $\mu_{q+1}$ due to $\gamma_1+\gamma_2\ne 0$ and  $\gamma_1\gamma_2\in\mu_{q+1}$. This together with \eqref{eq-x} one can conclude that any $x$ satisfying
$$x^{2q^2}=\frac{b+\gamma_1}{\gamma_1+\gamma_2}=\frac{b+\gamma_1}{\Tr^{4n}_n(b)(b^{q^2+1}+1)^{-q}}$$
is a solution of \eqref{diff-eq}, where $\gamma_1\in\mu_{q+1}$ is a solution of \eqref{eq-2nd}. Hence, when $\gamma_1\ne\gamma_2$ occurs, \eqref{diff-eq} has either no solution or two solutions if $b\not\in\mathbb{F}_{q^2}$.

Recall that $\omega_i=\#\{b\in\mathbb{F}_{2^{4n}}: \delta(1, b)=i\}$ and $\mathbb{S}=\{\omega_0,\omega_2,\cdots, \omega_{\delta_F}\}$ is the differential spectrum of $F(x)=x^d$ over $\mathbb{F}_{2^{4n}}=\mathbb{F}_{q^4}$. Then, based on the above discussions we can obtain that $\mathbb{S}=\{\omega_0,\omega_2,\omega_{q^2-q}, \omega_{q^2}\}$ with $\omega_{q^2-q}=q$ and $\omega_{q^2}=1$. Combining this with the well-known identities
\begin{eqnarray*}
 \sum_{i=0}^{q^4}\omega_i=q^4, \;\;\sum_{i=0}^{q^4}i\omega_i=q^4
\end{eqnarray*}
gives
\begin{eqnarray*}
\omega_0=(\frac{q^3}{2}-1)(q+1),\;
\omega_2=\frac{q^4-q^3}{2}.
\end{eqnarray*}
This completes the proof.


\end{document}